\documentclass[letterpaper,twocolumn,prl,aps,superscriptaddress,amsmath,amssymb,floatfix]{revtex4-1}

\usepackage{natbib}
\usepackage{mathptmx}
\usepackage[utf8]{inputenc}
\usepackage{color}
\usepackage{verbatim}
\usepackage{float}
\usepackage{amssymb,amsmath,amssymb,amsthm,mathrsfs}
\usepackage{graphicx}
\usepackage{esint}
\usepackage{siunitx}
\usepackage{soul}
\usepackage[unicode=true,
 bookmarks=true,bookmarksnumbered=false,bookmarksopen=false,
 breaklinks=false,pdfborder={0 0 1},backref=false,colorlinks=true]
 {hyperref}
\hypersetup{
 linkcolor=magenta,urlcolor=blue,citecolor=blue,pdfstartview={FitH},hyperfootnotes=false}
\usepackage[english]{babel}

\makeatletter

\usepackage{textcomp}
\usepackage{epstopdf}

\pdfpageheight\paperheight
\pdfpagewidth\paperwidth

\@ifundefined{textcolor}{}{%
 \definecolor{BLACK}{gray}{0}
 \definecolor{WHITE}{gray}{1}
 \definecolor{RED}{rgb}{1,0,0}
 \definecolor{GREEN}{rgb}{0,1,0}
 \definecolor{BLUE}{rgb}{0,0,1}
 \definecolor{CYAN}{cmyk}{1,0,0,0}
 \definecolor{MAGENTA}{cmyk}{0,1,0,0}
 \definecolor{YELLOW}{cmyk}{0,0,1,0}
}

\usepackage{xcolor}
\usepackage{soul}
\newcommand{\bra}[1]{\ensuremath{\left\langle#1\right|}}
\newcommand{\ket}[1]{\ensuremath{\left|#1\right\rangle}}

\definecolor{blue}{rgb}{0,0,1}
\definecolor{red}{rgb}{1,0,0}
\definecolor{green}{rgb}{0,1,0}

\usepackage{soul}

\makeatother

\begin{document}
\title{Optimal control with flag qubits}

\author{Liang-Xu Xie}
\thanks{These authors contribute equally to this work.}
\affiliation{Laboratory of Quantum Information, University of Science
and Technology of China, Hefei, Anhui 230026, China}

\author{Lui Zuccherelli de Paula}
\thanks{These authors contribute equally to this work.}
\affiliation{Laboratory of Quantum Information, University of Science
and Technology of China, Hefei, Anhui 230026, China}

\author{Weizhou Cai}
\affiliation{Laboratory of Quantum Information, University of Science
and Technology of China, Hefei, Anhui 230026, China}

\author{Qing-Xuan Jie}
\affiliation{Laboratory of Quantum Information, University of Science
and Technology of China, Hefei, Anhui 230026, China}

\author{Luyan Sun}
\affiliation{Center for Quantum Information, Institute for Interdisciplinary Information Sciences, Tsinghua University, Beijing 100084, China}
\affiliation{Hefei National Laboratory, Hefei 230088, China.}

\author{Chang-Ling Zou}
\affiliation{Laboratory of Quantum Information, University of Science
and Technology of China, Hefei, Anhui 230026, China}
\affiliation{CAS Center For Excellence in Quantum Information and Quantum Physics,
University of Science and Technology of China, Hefei, Anhui 230026, China}
\affiliation{Hefei National Laboratory, Hefei 230088, China.}

\author{Guang-Can Guo}
\affiliation{Laboratory of Quantum Information, University of Science
and Technology of China, Hefei, Anhui 230026, China}
\affiliation{CAS Center For Excellence in Quantum Information and Quantum Physics,
University of Science and Technology of China, Hefei, Anhui 230026, China}
\affiliation{Hefei National Laboratory, Hefei 230088, China.}

\author{Zi-Jie Chen}
\email{zijie@ustc.edu.cn}
\affiliation{Laboratory of Quantum Information, University of Science
and Technology of China, Hefei, Anhui 230026, China}

\author{Xu-Bo Zou}
\email{xbz@ustc.edu.cn}
\affiliation{Laboratory of Quantum Information, University of Science
and Technology of China, Hefei, Anhui 230026, China}
\affiliation{CAS Center For Excellence in Quantum Information and Quantum Physics,
University of Science and Technology of China, Hefei, Anhui 230026, China}
\affiliation{Hefei National Laboratory, Hefei 230088, China.}

\date{\today}
\begin{abstract}
High-fidelity quantum operations are the cornerstone of fault-tolerant quantum computation. In open quantum systems, traditional optimal control only passively resists decoherence, leaving environment-induced uncertainty as a fundamental performance bottleneck. To overcome this, we propose a new optimal control framework with flag ancillas and the Flag-GRAPE algorithm, which can actively tailor the system's noise structure. Through embedding post-selection directly into the objective function, Flag-GRAPE correlates decoherence errors with the ancilla's unexpected state. Subsequent measurement and post-selection effectively expel this uncertainty, circumventing the fidelity bounds of traditional control. Numerical simulations in a superconducting quantum circuit demonstrate a $51\%$ reduction in infidelity compared to traditional closed-system pulses and also show that such enhancement is robust across broad noise regimes. Furthermore, by actively converting unstructured decoherence into heralded erasure errors, Flag-GRAPE is inherently compatible with quantum error correction. We demonstrate this by initializing a logical cat-code state, showing that the combination between Flag-GRAPE and QEC yields immediate state preparation enhancements. This new framework can reduce hardware overhead for fault-tolerant architectures and open up a practical path toward logical state preparation gain in near-term experiments.

\end{abstract}

\maketitle 

\section{INTRODUCTION}

High-fidelity quantum state preparation and gate operations are foundational for realizing fault-tolerant quantum computation~\cite{NonClifford-ImprovedSimulationStabilizer2004,gottesman1998heisenberg,magicstate-2005PRA,eastinRestrictionsTransversalEncoded2009,bravyiMagicstateDistillationLow2012}. To approach the fidelity limits permitted by physical hardware, quantum optimal control techniques, particularly gradient-based pulse optimization algorithms such as gradient ascent pulse engineering (GRAPE)~\cite{closegrape2005}, became standard tools across mainstream platforms, such as superconducting circuits and trapped ions~\cite{dAlessandro2021QuantumControl,magannPulsesCircuitsBack2021}. However, a fundamental challenge limits the ultimate performance of these methods: realistic quantum systems inevitably couple with their environment, and the decoherence processes irreversibly inject uncertainty (\textit{i.e.}~entropy) into the system~\cite{Nielsen&Chuang,breuer-petruccioneOpenQuantumSystems2002,wiseman-milburnQuantumMeasurementControl2009}. The core mechanism of traditional optimal control relies on realizing strict or approximate unitary operations, which are inherently entropy-conserving~\cite{Nielsen&Chuang}. This indicates that no matter how exquisitely the control pulses are optimized, the achievable fidelity of a unitary is strictly bounded by the entropy produced by the environment. 
Breaking through this entropy {build-up} imposition remains a critical challenge in the field of quantum optimal control~\cite{kochControllingOpenQuantum2016}.

To address it, various strategies have been developed to mitigate environmental effects during the control process. Representative works include open-system GRAPE algorithms that incorporate the Lindblad master equation~\cite{chenOpenGRAPE,schulte-herbruggenOptimalControlGenerating2011, boutinResonatorResetCircuit2017a} or stochastic Schrödinger equations~\cite{abdelhafez2019gradient-based, goerzQuantumOptimalControl2022} into the optimization model, and control strategies that leverage machine learning techniques, such as reinforcement learning, to explore broader optimization landscapes~\cite{wuLearningRobustHighprecision2019}.
These approaches share a consistent core idea: acknowledging the presence of decoherence, they attempt to approach the optimal fidelity within the constraints of dissipative dynamics through more precise noise modeling or more sophisticated pulse design. Despite their respective merits and significant progress, they share the same fundamental limitation: the operational framework remains confined to implementing a strict or approximate unitary evolution. In essence, they passively adapt to the uncertainty imposed by decoherence, failing to provide a physical channel to actively extract this additional uncertainty from the system.

Drawing inspiration from the flag techniques utilized in quantum error correction (QEC)~\cite{chaoflagqubit2018,chamberlandFlagFaulttolerantError2018}, we propose a new optimal control framework based on a flag ancilla, and also develop an efficient algorithm named Flag-GRAPE, which offer a fundamentally new perspective on this problem. The core idea is to employ an auxiliary system to serve as a flag, and to directly embed the post-selection operation into the objective function for control optimization. This design fundamentally alters the nature of the control process, given that post-selection is a non-linear operation that can provide the system with a physical channel to extract entropy. During the optimization procedure, the algorithm tries to employ control pulses that guide the error components induced by decoherence into a specific subspace marked by the ancilla, such that they are subsequently suppressed via measurement and post-selection. Shifting from pure unitary optimization to measurement-conditioned process optimization, and sacrificing a portion of the operation's success probability, this mechanism yields a substantial enhancement in the fidelity of the target operation.

From another perspective, this approach is equivalent to actively tailoring the noise structure of the system through the control pulses. It converts unstructured decoherence errors into well-defined erasure errors on the system~\cite{grasslCodesQuantumErasure1997}. Most QEC protocols, such as surface code~\cite{fowlerSurfaceCodesPractical2012}, can realize a significantly higher fault-tolerance threshold for erasure errors compared to general depolarizing errors~\cite{erasureerror-ThresholdsTopologicalCodes2009, wu2022erasure-Rydbergatomarrays,kubicaErasureQubitsOvercoming2023}. Therefore, the quantum operation realized by the flag-based scheme with the Flag-GRAPE algorithm can show a strong compatibility with error-correction protocols.

We systematically validate the Flag-GRAPE algorithm through numerical simulations on a superconducting quantum circuit~\cite{blaisCircuitQuantumElectrodynamics2021, voolIntroductionQuantumElectromagnetic2017}. Under realistic open-system decoherence, this new algorithm can achieve a remarkable $51\%$ reduction in average infidelity compared to Closed-GRAPE~\cite{closegrape2005}, the conventional optimization approach based on closed-system dynamics. Furthermore, the method exhibits strong robustness against elevated noise strength, maintaining high preparation precision while relaxing hardware constraints. Finally, we further apply Flag-GRAPE to the initial state preparation with cat code. The corresponding numerical results demonstrate the algorithm's high compatibility with quantum error correction codes, showing a $50\%$ reduction in average infidelity compared to Closed-GRAPE.

The structure of this paper is organized as follows: in Sec.~\ref{sec theory}, we establish the theoretical framework of the flag-based scheme and Flag-GRAPE algorithm, where an ancillary system combined with post-selection measurements can effectively remove the uncertainty induced by decoherence. In Sec.~\ref{sec: numerical results}, we evaluate the Flag-GRAPE algorithm's performance under realistic open-system parameters, benchmarking it against Closed-GRAPE. In Sec.~\ref{sec break-even}, we further demonstrate that this new scheme is compatible with quantum error correction, showing the initial state preparation of a four-legged cat state as an example. Finally, conclusions and future research directions are summarized in Sec.~\ref{sec discussion}.

\section{Theory}\label{sec theory}

\begin{figure}[t]
    \centering
        \includegraphics[width=\columnwidth]{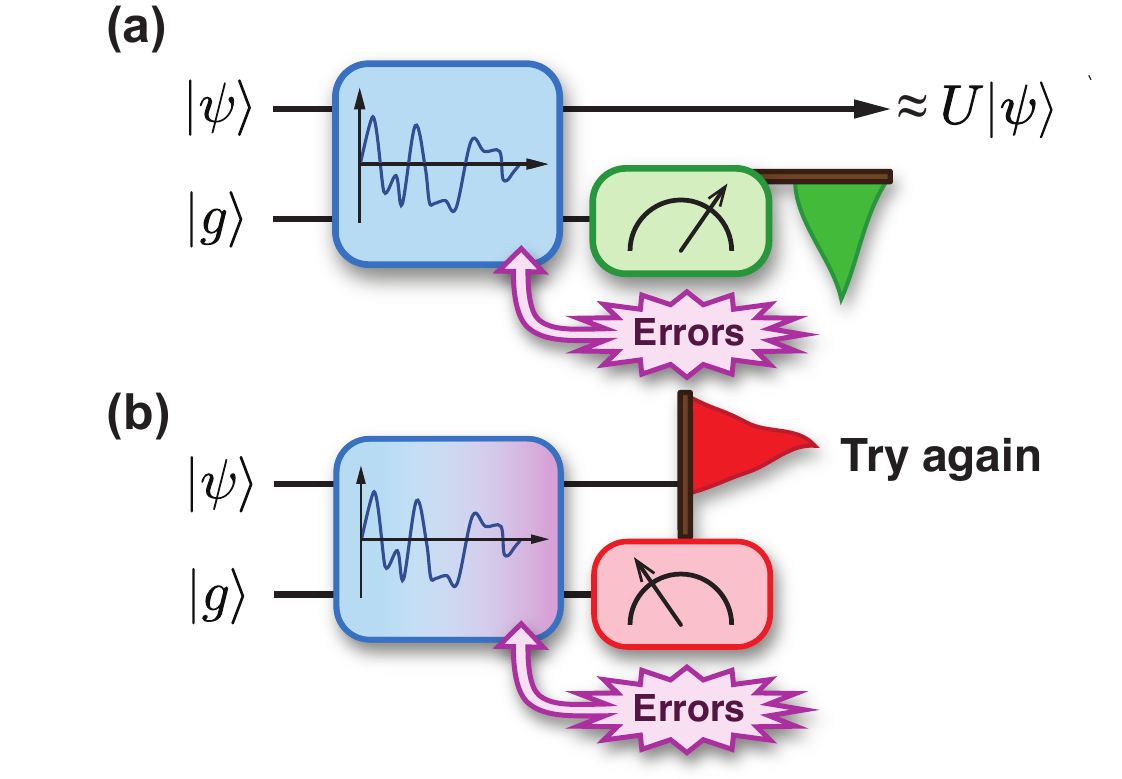} 
\caption{\label{Fig1}
\textbf{Schematic of the suppression of noise-induced uncertainty via a flag ancilla.} The target system inevitably suffers from environmental noise during gate operations. By introducing an auxiliary flag qubit, a post-selection measurement can be performed immediately after the target operation to evaluate the process. \textbf{(a) }If the ancilla is measured in the expected state, no error is flagged, and the resulting state of the target system is accepted. \textbf{(b) }Conversely, if the ancilla measurement yields an unexpected outcome  and the flag is raised, it deterministically indicates that an error has occurred during the evolution, and the corresponding corrupted state is rejected.
}
\end{figure}

As illustrated in Fig.~\ref{Fig1}, the physical setup for implementing a flag-based scheme consists of two components: a target system and an ancillary system. The objective of the scheme is to apply a high-fidelity quantum gate to the target system. In this architecture, the ancilla has a dual purpose: first, during the operation, it provides necessary nonlinearity or entangling capabilities to enable the control drives to steer the target system. Second, and more significant to our case, it acts as a repository for the uncertainty introduced by the composite system's decoherence. By performing a measurement on the ancilla, followed by a post-selection, this noise-induced uncertainty can be effectively stripped away from the target system's dynamics. 

The ancillary system is initialized in a specific state, \textit{e.g.}~the ground state $|g\rangle$, and our goal is to implement a high-fidelity unitary operation $U$ on the target system. During the gate operation, the continuous evolution of the composite open system in the presence of noise is governed by the Lindblad master equation~\cite{lindblad_generators_1976,goriniCompletelyPositiveDynamical1976}  ($\hbar=1$):
\begin{equation}\label{Eq3MasterEq}
\frac{d}{dt}\rho(t)=-i [H(t),\rho(t)]+ \mathscr{L}(\rho(t)),
\end{equation}
where $\rho(t)$ is the density matrix of the composite system, $H(t)$ is the total Hamiltonian, and $\mathscr{L}$ is the dissipator superoperator describing the irreversible contribution to the system's evolution:
\begin{equation}
\mathscr{L}(\rho(t))
=\sum_{\mu}\gamma_{\mu}\!\left(
c_{\mu}\rho(t) c_{\mu}^{\dagger}
-\frac{1}{2}c_{\mu}^{\dagger}c_{\mu}\rho(t)
-\frac{1}{2}\rho(t)c_{\mu}^{\dagger}c_{\mu}
\right),
\end{equation}
where $c_\mu$ are the jump operators with corresponding strengths $\gamma_\mu$. The final state after this continuous evolution is $\rho(T)$, where $T$ is the total gate duration of the control process. Subsequently, a projective measurement is performed on the ancillary system. The retained  quantum state after post-selection is:
\begin{equation}
\rho_f=\frac{M_0\rho(T) M_0}{p_0},
\end{equation} 
where $M_0$ is the projective operator onto the target state of the ancilla system and  $p_0$ is the success probability of the post-selection:
\begin{equation}\label{p_0}
p_0 = \text{Tr}[M_0\rho(T)M_0]=\text{Tr}[M_0\rho(T)].
\end{equation}

To implement the desired unitary operation $U$, the driving pulses of the system can be numerically optimized. GRAPE~\cite{closegrape2005} is an iterative gradient-based numerical optimization algorithm and is one of the most popular tools for this task. By employing a piecewise-constant control approach, the Hamiltonian is kept fixed over small time intervals, and the pulse amplitudes are hence optimized. More explicitly, the time-dependent Hamiltonian $H(t)$ can be expressed as:
\begin{equation}
    H(t) = H_0 + \sum_{j=1}^{J} u^{(j)}(t) H_c^{(j)},
\end{equation}
where $H_0$ is the drift Hamiltonian, $H_c^{(j)}$ is the $j$-th control Hamiltonian, and $u^{(j)}(t)$ represents the corresponding control amplitude.  $T$ is uniformly divided into $N$ equal steps of duration $\delta t=T/N$, during each of which the control Hamiltonian remains constant, \textit{i.e.},~$u^{(j)}(t_k)=u_k^{(j)} $ for $t_k$ inside the $k$-th step, and the corresponding piecewise-constant total Hamiltonian can be accordingly rewritten as $H_k=H_0 + \sum_{j=1}^{J} u^{(j)}_k H_c^{(j)}$.

In optimal control theory, a target operation is typically formulated as a set of state transfers $\{\ket{\psi^i_0}\rightarrow \ket{\psi^i_t}=U\ket{\psi^i_0}\}$, where $\ket{\psi^i_0}$ are the initial states, $\ket{\psi^i_t}$ the respective target states, and the superscript $i$ indexes the $i$-th constraint required to define the target operation. In the conventional scheme, the objective is to make the evolution driven by the control pulses approximate $U$ as closely as possible, where the conventional objective function $\Phi_{\text{conv}}$  to be minimized is the sum of infidelities $f_{\text{pre}}^i$ for all the final states $\rho^i(T)$:
\begin{equation}\label{infid conventional}
        f^i_{\text{pre}} = 1 - \langle\psi_t^i|\rho^i(T)|\psi_t^i\rangle,
        \qquad
        \Phi_{\text{conv}} = \sum_i f^i_\text{pre}.
\end{equation}
In contrast, the flag-based scheme proposed here effectively suppresses part of the system uncertainty through the measurement of the ancilla. Consequently, the objective function for this scheme is defined by the infidelity between the target state and the final state \textit{after} post-selection:
\begin{equation}\label{infid flag}
    f^i_{\text{post}} =  1 - \langle\psi_t^i|\rho_f^i|\psi_t^i\rangle = 1 - \frac{\langle\psi_t^i|\rho^i(T)|\psi_t^i\rangle}{p_0}.
\end{equation}
To realize the optimization of the operation after post-selection, $\Phi_\text{flag} = \sum_i f^i_{\text{post}}$ should be minimized, and this is the objective function of the Flag-GRAPE algorithm. 

To optimize the control parameters, we must evaluate the gradient of the objective function, given by the infidelity of the post-selected final state:
\begin{equation}
        \frac{\partial f^i_{\text{post}}}{\partial u_k^{(j)} }
            \!=\!-\frac{1}{p_0}
                 \frac{\partial \bra{\psi_t^i}\rho^i(T)\ket{\psi_t^i}}{\partial u_k^{(j)}}
            +   \frac{\bra{\psi_t^i}\rho^i(T)\ket{\psi_t^i}}{p_0^2}
                \frac{\partial p_0}{\partial u_k^{(j)}}.
\end{equation}

In an open quantum system, evaluating this exact gradient requires solving the full Lindblad master equation, which is computationally prohibitive for large-dimensional systems. Instead, we approximate the open system dynamics using quantum trajectories~\cite{abdelhafez2019gradient-based, goerzQuantumOptimalControl2022,schaferControlStochasticQuantum2021,propsonRobustQuantumOptimal2022}. This method uses a finite number $M$ of quantum trajectories to approximate the behavior of the the full analytical master equation in Eq.~\ref{Eq3MasterEq}. To further solve the computational challenge,  
we implement an accelerated trajectory method afforded by the following two key techniques:
\begin{enumerate}
    \item Efficient sampling: we pre-calculate the deterministic no-jump trajectory to extract the jump probabilities at each step following the method in Ref.~\cite{abdelhafez2019gradient-based}. Since the system's decoherence rates are relatively weak in the current physical platform, no-jump trajectories vastly dominate the distribution; therefore, such a pre-calculation can prevent waste of computational resources by repeatedly sampling the same trivial paths. Next, we consider the trajectories with one single jump, neglecting those with more because of the weak decoherence rates. From the pre-calculated no-jump trajectory, we extract the probability distribution that the jump will happen at a certain step $k$, and generate $M-1$ single-jump trajectories with the location of their jumps distributed accordingly. The system's density matrix is thus the \textit{weighted} average of the $M$ trajectory states.
    \item Efficient derivation: drawing inspiration from the conventional GRAPE algorithm in closed systems, the calculation of  the gradient can be realized by utilizing forward and backward propagation of the trajectory 's initial and target state. \textit{E.g.}, during the computation of $\partial p_0/\partial  u_k^{(j)}$, instead of having to numerically calculate the differential of the states at step $k$, terms like $\bra{\psi_0^i} \xi^\dagger_{1} \cdots \xi^\dagger_{N} M_0 \xi_{N} \cdots \xi_{k+1}H_c^{(j)}\xi_{k}\cdots \xi_{1} \ket{\psi_0^i}$ will appear, where $\xi_k$ are (non-unitary) Kraus operators~\cite{kraus_general_1971} that describe the evolution in the $k$-th step: if no-jump happens,  $\xi_k$ is a propagator with a non-Hermitian effective Hamiltonian; while, if there is a jump, $\xi_k$ is proportional to the respective jump operator.
\end{enumerate}

With these two techniques, the computational complexity of each trajectory is the same as that of the conventional, closed version of the GRAPE algorithm, making this numerical algorithm practical for large dimensional system. A more detailed discussion can be found in the Appendix.

\section{Numerical Simulation}\label{sec: numerical results}

\begin{figure*}
\centering
\includegraphics[scale=1]{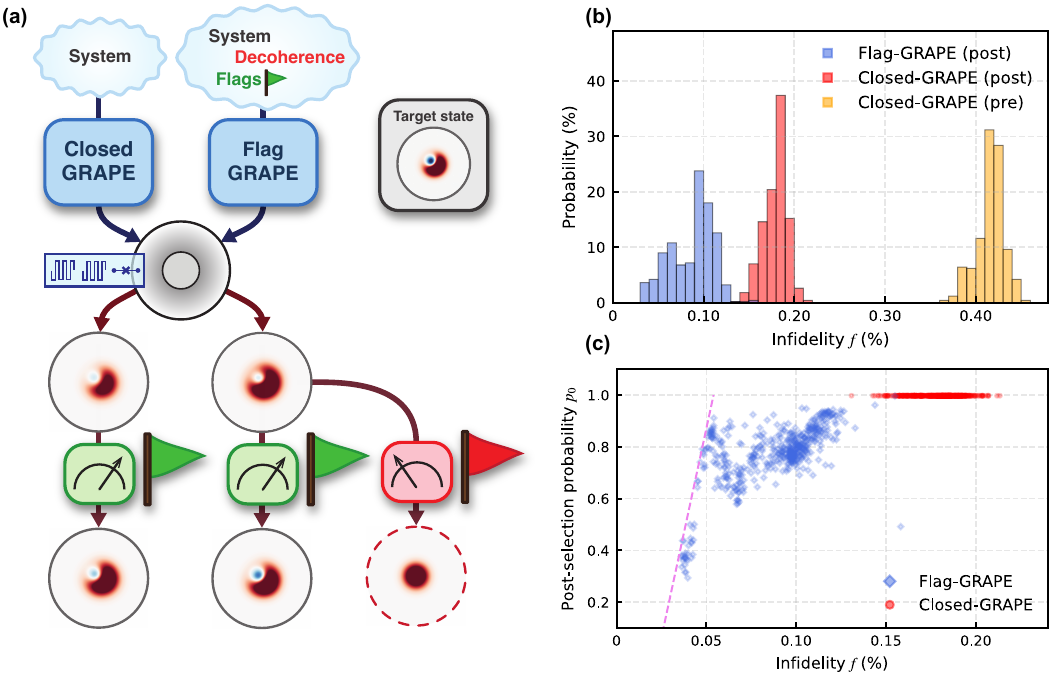}
\caption{\label{fig:Fig2} \textbf{Simulation results of the Flag-GRAPE algorithm}  
    (\textbf{a})~Schematic of the superconducting platform and the optimization workflows. Both Closed-GRAPE and Flag-GRAPE can suppress decoherence-induced uncertainty via measurement of the ancilla and post-selection. The flag-based scheme actively correlates decoherence errors with the ancilla's excited state, allowing noisy evolutions to be effectively flagged and discarded. The inset Wigner functions schematic illustratively compare the target state $\ket{0}+e^{-i\pi/4}\ket{1}$ prepared under these distinct strategies.
    (\textbf{b})~Infidelity distribution of 500 independently optimized Closed-GRAPE pulses and corresponding Flag-GRAPE pulses. Blue and red histograms denote the infidelities $f_\text{post}$ of Flag-GRAPE and Closed-GRAPE pulses with post-selection, respectively. The yellow histogram represents infidelities $f_\text{pre}$ from Closed-GRAPE pulses without post-selection. For visual clarity, the results for Flag-GRAPE pulses without post-selection, which distribute much further to the right, are shown in the Appendix.
    (\textbf{c})~Scatter plot illustrating the correlation between the final infidelity and the post-selection success probability for the post-selected Flag-GRAPE and Closed-GRAPE pulses shown in (b).
        }
\end{figure*}

To evaluate the performance of the Flag-GRAPE algorithm, we conduct numerical simulations based on a superconducting cavity system, one of the leading candidates for quantum information processing. As illustrated in Fig.~\ref{fig:Fig2}(a), this platform comprises a high-quality superconducting cavity for encoding and storing quantum states, dispersively coupled to an ancillary transmon qubit. Such a superconducting circuit is highly suitable for the flag-based scheme, since the transmon not only provides the requisite nonlinearity for universal operations, but also directly functions as the flag qubit to extract the uncertainty induced by decoherence. In the interaction picture, the drift Hamiltonian of the aforementioned composite system can be written as~\cite{blaisCircuitQuantumElectrodynamics2021}:
\begin{equation}
    H_0 = \frac{\chi}{2}a^{\dagger}a\sigma_z,
\end{equation}
where $a^{\dagger}$, $a$ are the creation and annihilation operators of the cavity, $\sigma_z$ is the Pauli-Z operator of the transmon qubit, and $\chi$ is the cross-Kerr coupling strength. Through controllable coherent microwave drives on both subsystems, it is possible to achieve universal control with the following control Hamiltonians~\cite{blaisCircuitQuantumElectrodynamics2021}:
\begin{equation}
    \begin{split}
        &H_c^{(1)}=\sigma_x ,\qquad  H_c^{(2)}=\sigma_y, \\
        &H_c^{(3)} = a+a^{\dagger},\quad H_c^{(4)} = i(a-a^{\dagger}),
    \end{split}
\end{equation}
where $\sigma_x$ and $\sigma_y$ are the Pauli-X and Pauli-Y operators of the transmon qubit, respectively.
The dominant decoherence errors of this system are considered in the simulation, which include the single-photon loss of the cavity $c_1=a$, spontaneous decay  $c_2=\sigma_-$ and dephasing $c_3 = \sigma_z$ of the ancilla.

Here, we consider the case with single-constraint condition as an illustrative example, which is equivalent to an initial state preparation task. The cavity and the transmon are initialized in vacuum and ground states, \textit{i.e.}, $  \ket{\psi_0}=\ket{0}\otimes\ket{g}$ . The goal of the operation is to initialize the target state 
\begin{equation}
    \ket{\psi_t} = \ket{0}+e^{-i\pi/4}\ket{1}
\end{equation}
in the cavity. After the implementation of the optimized driving pulses, noise suppression is realized by measuring the ancillary transmon qubit and the corresponding projective measurement is $M_0 = I \otimes |g\rangle\langle g|$.

Fig.~\ref{fig:Fig2}(b) shows the infidelity distribution of 500 independently optimized pulses generated by both the Closed-GRAPE, \textit{i.e.}~the conventional approach  based only on closed-system dynamics~\cite{closegrape2005}, and the proposed Flag-GRAPE algorithms. By simulating these pulses under the open-system master equation, we obtain their state preparation infidelities. The distribution reveals fidelity improvements in both algorithms after post-selection: the average infidelity of Closed-GRAPE is suppressed from $0.42\% \pm 0.02\% $ to $0.18\% \pm 0.01\%$ with post-selection, while Flag-GRAPE achieves a reduction from $23\%\pm 12\%$ to $0.088\% \pm 0.024\%$. This strongly confirms the hypothesis that measuring the ancillary qubit is an effective method of extracting uncertainty from the target system. On average, Flag-GRAPE achieves a  $51\%$ reduction in final infidelity $f_\text{post}$ when compared to Closed-GRAPE, demonstrating its superior efficacy. Furthermore, comparing the best-performing pulse from the 500 samples, Flag-GRAPE yields a best infidelity of $0.036\%$, compared to $0.13\%$ from Closed-GRAPE, marking a remarkable $72\%$ relative improvement.

Fig.~\ref{fig:Fig2}(c) further illustrates the correlation between the post-selection success probability $p_0$ and the infidelity $f$. Because Closed-GRAPE does not incorporate post-selection into its objective function, the results are tightly clustered near a probability of 1. In contrast, Flag-GRAPE pulses skew distinctly toward the bottom-left region of the plot. This indicates that the algorithm actively correlates gate noise with the ancilla's excited state, deliberately sacrificing a fraction of the success probability to filter out uncertainty via post-selection. Notably, the Flag-GRAPE data points exhibit a clear, sharp boundary on the left side (violet dashed line) roughly around $f\gtrapprox(2.3+3.1 \times p_0)\times 10^{-4}$. This heuristic line indicates the optimal physical trade-off between success probability and fidelity. Operating along this boundary allows the conversion of classical resources (\textit{e.g.}, repeated initializations or system redundancy) into quantum resources (lower infidelity), thereby enabling deeper quantum circuits. Extrapolating from this relation, the theoretical optimal infidelity that could be reached without post-selection (\textit{i.e.}~$p_0=1$) is approximately $5.4\times 10^{-4}$, and Flag-GRAPE successfully found a pulse $33\%$ lower than this limit.

To minimize computational overhead, the Flag-GRAPE algorithm is calculated using the results from Closed-GRAPE as initial pulses. The noisy system is simulated with realistic parameters, where cross-Kerr coupling strength is $\chi = 2.59$ MHz, cavity decay rate is $\gamma_1/2\pi = 0.275$ kHz, qubit relaxation rate is $\gamma_2/2\pi = 0.81$ kHz, and qubit dephasing rate is $\gamma_3/2\pi = 8.25$ kHz. The initial state evolves over a total gate duration of $T = 0.1~\mu\text{s}$ , which is equally divided into $N = 1000$ uniform time steps. For the trajectory sampling in Flag-GRAPE, we utilize $M = 50$ trajectories per iteration. Numerical verifications confirm that this sample size provides sufficiently accurate gradients for stable convergence (see Appendix for more details).

\begin{figure}
\centering
\includegraphics[width=\columnwidth]{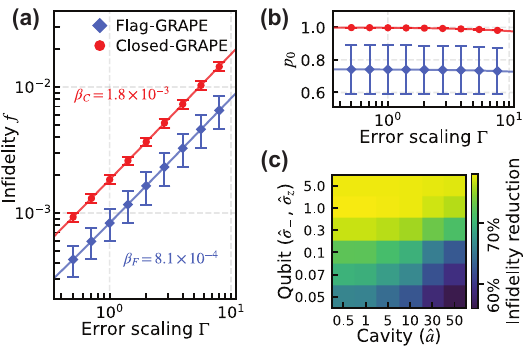}
\caption{\label{fig:Fig3} 
    \textbf{Performance of the algorithms with varying error rate.}
    \textbf{(a)}~Comparison of average infidelity of $10$ optimized pulses with Flag-GRAPE (blue) and Closed-GRAPE (red) where all error rates are multiplied equally by an error scaling factor $\Gamma$.
    The curves are fitted linearly with the respective prefactor $\beta$.
    \textbf{(b)}~Average post-selection probability $p_0$ for the same set.
    \textbf{(c)}~Reduction in the best infidelity by Flag-GRAPE pulses compared with Closed-GRAPE pulse for different error scalings in the cavity (decay $\hat a$) and qubit (decay $\sigma_-$ and dephasing $\sigma_z$, varied together).
    }
\end{figure}

To explore the robustness of the flag-based scheme against varying noise intensities and assess the system's sensitivity to different noise channels, we randomly selected 10 Closed-GRAPE pulses and 10 Flag-GRAPE pulses from the dataset presented in Fig.~\ref{fig:Fig2}(b) for further numerical simulations.
Fig.~\ref{fig:Fig3}(a) illustrates the infidelity trends under different noise strengths. Here, the noise intensity is parameterized as $\{\Gamma\times\gamma_i\}$, where $\Gamma$ is a scaling factor, with $\Gamma = 1$ corresponding to the baseline parameters in Fig.~\ref{fig:Fig2}(b). Flag-GRAPE maintains a consistent advantage across a broad parameter regime, and the infidelity exhibits a linear dependence on $\Gamma$. Based on a linear fit,  $f = \beta\times \Gamma + \text{const}$, we extract the scaling coefficients $\beta_C=1.8\times10^{-3}$ and $\beta_F=8.1\times10^{-4}$ for Closed-GRAPE and Flag-GRAPE, respectively, with negligibly small constant terms (less than $10^{-5}$) . Furthermore, Fig.~\ref{fig:Fig3}(b) demonstrates that the post-selection success probability remains almost unchanged with respect to $\Gamma$, indicating a highly stable yield even under different noise strengths.

Fig.~\ref{fig:Fig3}(c) shows the relative improvement of Flag-GRAPE over Closed-GRAPE, defined as $1-f_{\text{post}}^\text{(Flag)}/f_{\text{post}}^\text{(Closed)} $, where $f_{\text{post}}$ is the infidelity after post-selection (Eq.~\ref{infid flag}), by independently scaling the cavity photon loss and the combined ancilla noise (incorporating both relaxation and dephasing equally). The results reveal that, when ancilla errors dominate the open dynamics, the optimized pulses achieve an improvement of nearly $75\%$. Conversely, if cavity spontaneous emission is the dominant decoherence mechanism, this enhancement is restricted to approximately $60\%$. This indicates that the current superconducting architecture is fundamentally more sensitive to ancilla-induced errors , showing a clear direction for targeted experimental improvements. Furthermore, our results demonstrate that the flag-based scheme is highly robust against variations in decoherence strength, maintaining a significant improvement over Closed-GRAPE throughout the whole range.

\section{compatibility with quantum error correction}\label{sec break-even}

Beyond mitigating physical noise at the hardware level, the {flag-based} scheme also shows a natural compatibility with quantum error correction (QEC) architectures. First, fault-tolerant quantum computation fundamentally requires high-fidelity state preparation to initialize logical encodings, or to execute universal logical gates via magic state preparation, distillation, and injection. Second, and more significant, the flag-based scheme actively tailors the operational noise structure, effectively converting unstructured decoherence into heralded erasure errors~\cite{kubicaErasureQubitsOvercoming2023,grasslCodesQuantumErasure1997}. This conversion is profoundly advantageous: popular QEC codes, such as the surface code, exhibit an erasure error threshold approaching $50\%$~\cite{erasureerror-ThresholdsTopologicalCodes2009}, vastly outperforming their $1\%$ threshold for standard depolarizing noise~\cite{fowlerSurfaceCodesPractical2012}. Consequently, utilizing Flag-GRAPE for state initialization and active noise tailoring can drastically reduce the resource overhead demanded by top-level QEC protocols.

To better illustrate the compatibility with QEC, the objective function must be further generalized. In the situation encoding only a single logical qubit, the infidelity in Eq.~\ref{infid flag} can be rewritten using Pauli operator expectations: 
\begin{align}\label{logical fidelity}
   f^i_{\text{post}} &= 1 - \langle\psi_t^i|\rho^i_f|\psi_t^i\rangle \\
    &   = \frac{1}{2}-\frac{1}{2} \big(\epsilon_X  \text{Tr}[\rho^i_f X_L]+\epsilon_Y  \text{Tr}[\rho^i_f Y_L]+\epsilon_Z  \text{Tr}[\rho^i_f Z_L]\big)  \notag
\end{align}
where $\epsilon_j =\langle\psi_t^i|\sigma_j|\psi_t^i\rangle$ ($j=X,Y,Z$) are the coefficients of the target state, and $\{\text{Tr}[\rho_f X_L],\text{Tr}[\rho_f Y_L],\text{Tr}[\rho_f Z_L]\}$ denote the measured Pauli outcomes. The gradient evaluation method, previously established  in Sec.~\ref{sec theory}, remains directly applicable. The primary advantage of this reformulation is that it enables gradient computation directly through logical-level quantum state tomography.

However, logical-level quantum state tomography is more complicated for a QEC code. Unlike unencoded systems, QEC codes utilize redundant Hilbert spaces to protect quantum information, and this redundancy cannot be neglected. As an illustrative example, in a simple repetition code, both the primary code space (\textit{e.g.}, $\{|000\rangle, |111\rangle\}$) and subsequent error subspaces (\textit{e.g.}, $\{|100\rangle, |011\rangle\}$ after a bit-flip) are valid, recoverable spaces. Therefore, if we want to calculate the infidelity, we need to decode the final state via syndrome measurements to identify its specific error subspace, and, subsequently, incorporate that subspace in the evaluation of the value of Pauli operators. As a concrete example, using the same repetition code, after error syndrome, the state $\rho$ in the primary code space becomes $S_0\rho S_0$, while in the error subspaces it becomes $S_1\rho S_1$, for syndromes $S_0=\ket{000}\bra{000}+\ket{111}\bra{111}$ and $S_1=\ket{100}\bra{100}+\ket{011}\bra{011}$. The Pauli Z operator in these two subspaces are $Z_0=\ket{000}\bra{000}-\ket{111}\bra{111}$ and $Z_1=\ket{100}\bra{100}-\ket{011}\bra{011}$, respectively. Therefore, the \textit{value} of the Pauli $Z$ operator is $\text{Tr}[Z_0 S_0 \rho S_0]+\text{Tr}[Z_1 S_1 \rho S_1]=\text{Tr}[(S_0 Z_0 S_0+S_1 Z_1 S_1) \rho]$. This is equivalent to measuring the value of the \textit{logical}-$Z$ operator: $\text{Tr}[Z_L  \rho]$, where $Z_L=S_0 Z_0 S_0+S_1 Z_1 S_1=Z_0+Z_1$.

Equivalent Pauli operators $X_{L},Y_{L},Z_{L} $ can be constructed for any other encoding protocol.
Thus Eq.~\ref{logical fidelity} becomes the logical infidelity after decoding and its gradients can be computed via the algorithm described in section~\ref{sec theory}, because the values that need calculation in Eq.~\ref{logical fidelity} are $\text{Tr}[\rho^i_f X_L],\text{Tr}[\rho^i_f Y_L]$ and $\text{Tr}[\rho^i_f Z_L]$, whose formulas follow the same structure as $p_0$ in Eq.~\ref{p_0}.
Therefore, the control pulses can be optimized within the flag-based framework. In summary, we can realize the state initialization of a QEC code with redundant Hilbert spaces, which is a powerful tool for QEC.

The framework above is generalizable for any quantum encoding. In particular, it can also be applied to the four-component cat code, where the logical basis is encoded as :
\begin{equation}\label{cat code}
    \begin{split}
            &\ket{0_L}\propto \big( \ket{\alpha}+\ket{-\alpha}+\ket{i\alpha}+\ket{-i\alpha}\big) ,\\
            &\ket{1_L}\propto \big( \ket{\alpha}+\ket{-\alpha}-\ket{i\alpha}-\ket{-i\alpha}\big),
    \end{split}
\end{equation}
and the objective is to prepare the target encoded state
\begin{equation}\label{target_state}
    \ket{\bar \psi_t}= \ket{0_L}+e^{-i\pi/4}\ket{1_L}.
\end{equation}
Here, the logical $|0_L\rangle$ and $|1_L\rangle$ states possess distinct photon-number parities, residing in defined subspaces $\sum_n\ket{4n}\bra{4n}$ and $\sum_n\ket{4n+2}\bra{4n+2}$. Because a single-photon loss merely transitions the system into another recoverable parity subspace, the odd subspace remains an acceptable subspace for QEC. The corresponding equivalent logical tomography operator for Pauli-Z measurement is $Z_\text{cat} =  Z^+-Z^-$ where
\begin{align}
    Z^+ &= \sum_n\big(\ket{4n}\!\bra{4n} + \ket{4n+3}\!\bra{4n+3}\big),\\
    Z^- &= \sum_n\big(\ket{4n+1}\!\bra{4n+1} +\ket{4n+2}\!\bra{4n+2}\big).
\end{align}
Following the method in~\cite{chen2026faulttolerantpreparationarbitrarylogical}, we can then have the logical tomography operator $X_\text{cat} =2  X^+-I,$  where $ X^+ =  X_1 + X_2$ and 
\begin{align}
        X_1  &= D^\dagger(\alpha)\!\sum_{n=0,1}\ket n\! \bra n D(\alpha),\\
        X_2 &= (I\!-\!X_1)D^\dagger(-\alpha)\!\sum_{n=0,1}\ket n\! \bra n D(-\alpha)(I\!-\!X_1).
\end{align}
The corresponding Pauli-Y measurement can be constructed as $\quad Y_\text{cat} =  e^{-i\pi Z_\text{cat}/4} X_\text{cat} e^{i\pi Z_\text{cat}/4}$. More details can be found in the Appendix. These operators allow Flag-GRAPE to accurately evaluate the gradient and maximize the encoded logical fidelity. One must keep in mind that the projective measurement here still is $M_0 = I \otimes |g\rangle\langle g|$.

\begin{figure}[t]
    \centering
    \includegraphics[width=\columnwidth]{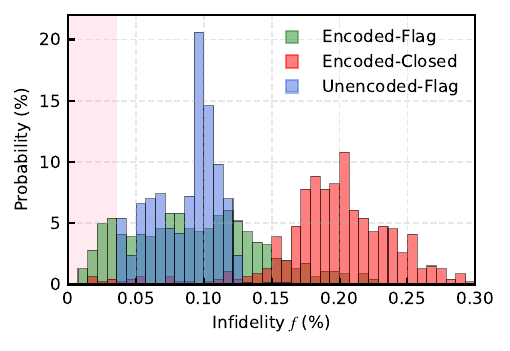}
\caption{\label{Fig_breakeven}\textbf{Logical state preparation of the cat-code within the flag-based scheme.}
    Probability distributions of infidelities for logical state $\ket{\bar \psi_t}= \ket{0_L}+e^{-i\pi/4}\ket{1_L}$ initialization with post-selection. The data is derived from 500 independently optimized Closed-GRAPE pulses (red histogram) and their corresponding Flag-GRAPE pulses (green histogram). For comparison, the blue histogram shows the result of unencoded state from Flag-GRAPE presented in Fig.~\ref{fig:Fig2}(b). The pink shaded region highlights the critical regime where the encoded infidelity surpasses the absolute best value ($\Phi<0.036\%$) achieved by the unencoded state. The gate duration for the encoded state is set to $0.2$ $\mu s$, and this value is determined via preliminary time optimizations as the near-shortest feasible time (see Appendix for details). All other system parameters remain the same as those from Fig.~\ref{fig:Fig2}(b).
    }
\end{figure}

To validate the compatibility of the flag-based scheme with the cat code, we evaluate its performance against the Closed-GRAPE algorithm. For the numerical simulation, 500 pulses were generated for the target state $\ket{{\bar \psi _t}}$  in Eq.~\ref{target_state} with Closed-GRAPE. After that, these were used as initial pulses for further optimization with  Flag-GRAPE based on the logical fidelity after decoding in Eq.~\ref{logical fidelity}.
As depicted in Fig.~\ref{Fig_breakeven}, Flag-GRAPE (green histogram) still outperforms the Closed-GRAPE algorithm (red histogram), delivering a $50\%$ reduction in average infidelity and a $46\%$ improvement for the best-performing pulse. For a comprehensive baseline, the unencoded Fock-state results from Fig.~\ref{fig:Fig2} (b) are also included in the distribution (blue histogram). Remarkably, about $13.4\%$ of the encoded Flag-GRAPE pulses achieve an infidelity lower than the best unencoded pulse. This indicates that an immediate state preparation enhancement may be attainable through the combination of cat code and Flag-GRAPE within the current parameter regime. Ultimately, these results demonstrate that the flag-based scheme is fundamentally compatible with QEC architectures, offering a highly effective tool for elevating fault-tolerant performance.

\section{Discussion}\label{sec discussion}
In summary, we introduce a new framework based on flag ancillas for quantum control in open systems and propose the Flag-GRAPE algorithm for efficient numerical optimization. Traditional optimal control strategies passively resist decoherence and only focus on increasing the fidelity, while our scheme also actively tailors the noise structure of the system during pulse optimization. Moreover, by deliberately correlating the uncertainty induced by decoherence with the excited state of the ancilla, and subsequently discarding this uncertainty via ancilla measurement and post-selection, our algorithm successfully circumvents the fidelity upper bound of purely unitary evolution.

Numerical simulations within a superconducting circuit show that Flag-GRAPE achieves a nearly $51\%$ reduction in infidelity in the state preparation task when compared to conventional, closed-system optimized pulses. Furthermore, the algorithm exhibits exceptional robustness across a wide range of noise strength. 

Crucially, the flag-based scheme and Flag-GRAPE algorithm are fundamentally compatible with quantum error correction architectures. It not only provides the high-fidelity initial states essential for fault-tolerant computation, but also actively transforms the dominant unstructured decoherence into heralded erasure errors. This low-level noise tailoring can effectively reduce the physical resource overhead required for top-level QEC. Further simulations based on the cat code indicate that, under near-term experimental parameters, integrating Flag-GRAPE with cat code can offer the opportunity to achieve an absolute fidelity gain in logical state preparation, demonstrating its practical value in the transition toward early fault-tolerant quantum computation.

Looking forward, we anticipate the direct experimental demonstration of the Flag-GRAPE algorithm on near-term superconducting quantum circuits or other hardware platforms such as trapped ions~\cite{paulElectromagneticTrapsCharged1990,leibfriedSingletrappedions2003}. For theoretical extensions, future work will extend this framework to actively tailor other noise structure including biased noise~\cite{aliferisFTagainstbiasednoise2008,tuckettbiased2018,bonillaataidesXZZXSurfaceCode2021}. This framework can also be utilized to synthesize more complicated high-fidelity logical gates and to prepare magic states. Furthermore, from a system-architecture perspective, future studies should quantitatively evaluate the performance of the top-level QEC code when the noise structure is tailored with Flag-GRAPE ,including the influences to the fault-tolerant threshold. 
Concurrently, further exploring the optimal boundary between the post-selection success probability and the ultimate fidelity limit will provide vital guidance for maximizing the conversion efficiency between classical and quantum resources.

\begin{acknowledgments}
\noindent This work was funded by the National Natural Science Foundation of China (Grant Nos. 12547179, 92265210, 12550006, 92365301, 92565301, 92165209, 12574539, 12404567), the Quantum Science and Technology-National Science and Technology Major Project (2021ZD0300200). This work is also supported by the Fundamental Research Funds for the Central Universities, the USTC Research Funds of the Double First-Class Initiative, the supercomputing system in the Supercomputing Center of USTC, and the USTC Center for Micro and Nanoscale Research and Fabrication.
\end{acknowledgments}



\smallskip{}
\bibliographystyle{Zou}
\bibliography{mycite}

\begin{thebibliography}{38}%
\makeatletter
\providecommand \@ifxundefined [1]{%
 \@ifx{#1\undefined}
}%
\providecommand \@ifnum [1]{%
 \ifnum #1\expandafter \@firstoftwo
 \else \expandafter \@secondoftwo
 \fi
}%
\providecommand \@ifx [1]{%
 \ifx #1\expandafter \@firstoftwo
 \else \expandafter \@secondoftwo
 \fi
}%
\providecommand \natexlab [1]{#1}%
\providecommand \enquote  [1]{``#1''}%
\providecommand \bibnamefont  [1]{#1}%
\providecommand \bibfnamefont [1]{#1}%
\providecommand \citenamefont [1]{#1}%
\providecommand \href@noop [0]{\@secondoftwo}%
\providecommand \href [0]{\begingroup \@sanitize@url \@href}%
\providecommand \@href[1]{\@@startlink{#1}\@@href}%
\providecommand \@@href[1]{\endgroup#1\@@endlink}%
\providecommand \@sanitize@url [0]{\catcode `\\12\catcode `\$12\catcode `\&12\catcode `\#12\catcode `\^12\catcode `\_12\catcode `\%12\relax}%
\providecommand \@@startlink[1]{}%
\providecommand \@@endlink[0]{}%
\providecommand \url  [0]{\begingroup\@sanitize@url \@url }%
\providecommand \@url [1]{\endgroup\@href {#1}{\urlprefix }}%
\providecommand \urlprefix  [0]{URL }%
\providecommand \Eprint [0]{\href }%
\providecommand \doibase [0]{http://dx.doi.org/}%
\providecommand \selectlanguage [0]{\@gobble}%
\providecommand \bibinfo  [0]{\@secondoftwo}%
\providecommand \bibfield  [0]{\@secondoftwo}%
\providecommand \translation [1]{[#1]}%
\providecommand \BibitemOpen [0]{}%
\providecommand \bibitemStop [0]{}%
\providecommand \bibitemNoStop [0]{.\EOS\space}%
\providecommand \EOS [0]{\spacefactor3000\relax}%
\providecommand \BibitemShut  [1]{\csname bibitem#1\endcsname}%
\let\auto@bib@innerbib\@empty
\bibitem [{\citenamefont {Aaronson}\ and\ \citenamefont {Gottesman}(2004)}]{NonClifford-ImprovedSimulationStabilizer2004}%
  \BibitemOpen
  \bibfield  {author} {\bibinfo {author} {\bibfnamefont {S.}~\bibnamefont {Aaronson}}\ and\ \bibinfo {author} {\bibfnamefont {D.}~\bibnamefont {Gottesman}},\ }\bibfield  {title} {\enquote {\bibinfo {title} {Improved simulation of stabilizer circuits},}\ }\href {\doibase 10.1103/PhysRevA.70.052328} {\bibfield  {journal} {\bibinfo  {journal} {Phys. Rev. A}\ }\textbf {\bibinfo {volume} {70}},\ \bibinfo {pages} {052328} (\bibinfo {year} {2004})}\BibitemShut {NoStop}%
\bibitem [{\citenamefont {Gottesman}(1998)}]{gottesman1998heisenberg}%
  \BibitemOpen
  \bibfield  {author} {\bibinfo {author} {\bibfnamefont {D.}~\bibnamefont {Gottesman}},\ }\bibfield  {title} {\enquote {\bibinfo {title} {The heisenberg representation of quantum computers},}\ }\href {\doibase 10.48550/arXiv.quant-ph/9807006} {\bibfield  {journal} {\bibinfo  {journal} {arXiv:quant-ph/9807006}\ } (\bibinfo {year} {1998}),\ 10.48550/arXiv.quant-ph/9807006}\BibitemShut {NoStop}%
\bibitem [{\citenamefont {Magnard}\ \emph {et~al.}(2005)\citenamefont {Magnard}, \citenamefont {Kurpiers}, \citenamefont {Royer}, \citenamefont {Walter}, \citenamefont {Besse}, \citenamefont {Gasparinetti}, \citenamefont {Pechal}, \citenamefont {Heinsoo}, \citenamefont {Storz}, \citenamefont {Blais},\ and\ \citenamefont {Wallraff}}]{magicstate-2005PRA}%
  \BibitemOpen
  \bibfield  {author} {\bibinfo {author} {\bibfnamefont {P.}~\bibnamefont {Magnard}}, \bibinfo {author} {\bibfnamefont {P.}~\bibnamefont {Kurpiers}}, \bibinfo {author} {\bibfnamefont {B.}~\bibnamefont {Royer}}, \bibinfo {author} {\bibfnamefont {T.}~\bibnamefont {Walter}}, \bibinfo {author} {\bibfnamefont {J.~C.}\ \bibnamefont {Besse}}, \bibinfo {author} {\bibfnamefont {S.}~\bibnamefont {Gasparinetti}}, \bibinfo {author} {\bibfnamefont {M.}~\bibnamefont {Pechal}}, \bibinfo {author} {\bibfnamefont {J.}~\bibnamefont {Heinsoo}}, \bibinfo {author} {\bibfnamefont {S.}~\bibnamefont {Storz}}, \bibinfo {author} {\bibfnamefont {A.}~\bibnamefont {Blais}}, \ and\ \bibinfo {author} {\bibfnamefont {A.}~\bibnamefont {Wallraff}},\ }\bibfield  {title} {\enquote {\bibinfo {title} {{Universal quantum computation with ideal Clifford gates and noisy ancillas}},}\ }\href {\doibase 10.1103/PhysRevA.71.022316} {\bibfield  {journal} {\bibinfo  {journal} {Phys. Rev. A}\ }\textbf {\bibinfo {volume} {71}},\ \bibinfo {pages} {22316}
  (\bibinfo {year} {2005})}\BibitemShut {NoStop}%
\bibitem [{\citenamefont {Eastin}\ and\ \citenamefont {Knill}(2009)}]{eastinRestrictionsTransversalEncoded2009}%
  \BibitemOpen
  \bibfield  {author} {\bibinfo {author} {\bibfnamefont {B.}~\bibnamefont {Eastin}}\ and\ \bibinfo {author} {\bibfnamefont {E.}~\bibnamefont {Knill}},\ }\bibfield  {title} {\enquote {\bibinfo {title} {Restrictions on {{Transversal Encoded Quantum Gate Sets}}},}\ }\href {\doibase 10.1103/PhysRevLett.102.110502} {\bibfield  {journal} {\bibinfo  {journal} {Phys. Rev. Lett.}\ }\textbf {\bibinfo {volume} {102}},\ \bibinfo {pages} {110502} (\bibinfo {year} {2009})}\BibitemShut {NoStop}%
\bibitem [{\citenamefont {Bravyi}\ and\ \citenamefont {Haah}(2012)}]{bravyiMagicstateDistillationLow2012}%
  \BibitemOpen
  \bibfield  {author} {\bibinfo {author} {\bibfnamefont {S.}~\bibnamefont {Bravyi}}\ and\ \bibinfo {author} {\bibfnamefont {J.}~\bibnamefont {Haah}},\ }\bibfield  {title} {\enquote {\bibinfo {title} {Magic-state distillation with low overhead},}\ }\href {\doibase 10.1103/PhysRevA.86.052329} {\bibfield  {journal} {\bibinfo  {journal} {Phys. Rev. A}\ }\textbf {\bibinfo {volume} {86}},\ \bibinfo {pages} {052329} (\bibinfo {year} {2012})}\BibitemShut {NoStop}%
\bibitem [{\citenamefont {Khaneja}\ \emph {et~al.}(2005)\citenamefont {Khaneja}, \citenamefont {Reiss}, \citenamefont {Kehlet}, \citenamefont {Schulte-Herbr{\"{u}}ggen},\ and\ \citenamefont {Glaser}}]{closegrape2005}%
  \BibitemOpen
  \bibfield  {author} {\bibinfo {author} {\bibfnamefont {N.}~\bibnamefont {Khaneja}}, \bibinfo {author} {\bibfnamefont {T.}~\bibnamefont {Reiss}}, \bibinfo {author} {\bibfnamefont {C.}~\bibnamefont {Kehlet}}, \bibinfo {author} {\bibfnamefont {T.}~\bibnamefont {Schulte-Herbr{\"{u}}ggen}}, \ and\ \bibinfo {author} {\bibfnamefont {S.~J.}\ \bibnamefont {Glaser}},\ }\bibfield  {title} {\enquote {\bibinfo {title} {{Optimal control of coupled spin dynamics: Design of NMR pulse sequences by gradient ascent algorithms}},}\ }\href {\doibase 10.1016/j.jmr.2004.11.004} {\bibfield  {journal} {\bibinfo  {journal} {J. Magn. Reson.}\ }\textbf {\bibinfo {volume} {172}},\ \bibinfo {pages} {296} (\bibinfo {year} {2005})}\BibitemShut {NoStop}%
\bibitem [{\citenamefont {D’Alessandro}(2021)}]{dAlessandro2021QuantumControl}%
  \BibitemOpen
  \bibfield  {author} {\bibinfo {author} {\bibfnamefont {D.}~\bibnamefont {D’Alessandro}},\ }\href {https://books.google.com.hk/books?id=n6A4EAAAQBAJ} {\emph {\bibinfo {title} {Introduction to Quantum Control and Dynamics}}},\ Advances in Applied Mathematics\ (\bibinfo  {publisher} {CRC Press},\ \bibinfo {year} {2021})\BibitemShut {NoStop}%
\bibitem [{\citenamefont {Magann}\ \emph {et~al.}(2021)\citenamefont {Magann}, \citenamefont {Arenz}, \citenamefont {Grace}, \citenamefont {Ho}, \citenamefont {Kosut}, \citenamefont {McClean}, \citenamefont {Rabitz},\ and\ \citenamefont {Sarovar}}]{magannPulsesCircuitsBack2021}%
  \BibitemOpen
  \bibfield  {author} {\bibinfo {author} {\bibfnamefont {A.~B.}\ \bibnamefont {Magann}}, \bibinfo {author} {\bibfnamefont {C.}~\bibnamefont {Arenz}}, \bibinfo {author} {\bibfnamefont {M.~D.}\ \bibnamefont {Grace}}, \bibinfo {author} {\bibfnamefont {T.-S.}\ \bibnamefont {Ho}}, \bibinfo {author} {\bibfnamefont {R.~L.}\ \bibnamefont {Kosut}}, \bibinfo {author} {\bibfnamefont {J.~R.}\ \bibnamefont {McClean}}, \bibinfo {author} {\bibfnamefont {H.~A.}\ \bibnamefont {Rabitz}}, \ and\ \bibinfo {author} {\bibfnamefont {M.}~\bibnamefont {Sarovar}},\ }\bibfield  {title} {\enquote {\bibinfo {title} {From {{Pulses}} to {{Circuits}} and {{Back Again}}: {{A Quantum Optimal Control Perspective}} on {{Variational Quantum Algorithms}}},}\ }\href {\doibase 10.1103/PRXQuantum.2.010101} {\bibfield  {journal} {\bibinfo  {journal} {PRX Quantum}\ }\textbf {\bibinfo {volume} {2}},\ \bibinfo {pages} {010101} (\bibinfo {year} {2021})}\BibitemShut {NoStop}%
\bibitem [{\citenamefont {{Nielsen, Michael A and Chuang}}(2010)}]{Nielsen&Chuang}%
  \BibitemOpen
  \bibfield  {author} {\bibinfo {author} {\bibfnamefont {I.~L.}\ \bibnamefont {{Nielsen, Michael A and Chuang}}},\ }\href {\doibase 10.1214/11-STS378} {\emph {\bibinfo {title} {{Quantum computation and quantum information}}}}\ (\bibinfo  {publisher} {Cambridge university press},\ \bibinfo {year} {2010})\BibitemShut {NoStop}%
\bibitem [{\citenamefont {Breuer}\ and\ \citenamefont {Petruccione}(2002)}]{breuer-petruccioneOpenQuantumSystems2002}%
  \BibitemOpen
  \bibfield  {author} {\bibinfo {author} {\bibfnamefont {H.-P.}\ \bibnamefont {Breuer}}\ and\ \bibinfo {author} {\bibfnamefont {F.}~\bibnamefont {Petruccione}},\ }\href {\doibase https://doi.org/10.1093/acprof:oso/9780199213900.001.0001} {{\selectlanguage {english}\emph {\bibinfo {title} {The {Theory} of {Open} {Quantum} {Systems}}}}}\ (\bibinfo  {publisher} {Oxford University Press},\ \bibinfo {year} {2002})\BibitemShut {NoStop}%
\bibitem [{\citenamefont {Wiseman}\ and\ \citenamefont {Milburn}(2009)}]{wiseman-milburnQuantumMeasurementControl2009}%
  \BibitemOpen
  \bibfield  {author} {\bibinfo {author} {\bibfnamefont {H.~M.}\ \bibnamefont {Wiseman}}\ and\ \bibinfo {author} {\bibfnamefont {G.~J.}\ \bibnamefont {Milburn}},\ }\href {\doibase https://doi.org/10.1017/CBO9780511813948} {{\selectlanguage {english}\emph {\bibinfo {title} {Quantum {Measurement} and {Control}}}}}\ (\bibinfo  {publisher} {Cambridge University Press},\ \bibinfo {address} {Cambridge},\ \bibinfo {year} {2009})\BibitemShut {NoStop}%
\bibitem [{\citenamefont {Koch}(2016)}]{kochControllingOpenQuantum2016}%
  \BibitemOpen
  \bibfield  {author} {\bibinfo {author} {\bibfnamefont {C.~P.}\ \bibnamefont {Koch}},\ }\bibfield  {title} {\enquote {\bibinfo {title} {Controlling open quantum systems: Tools, achievements, and limitations},}\ }\href {\doibase 10.1088/0953-8984/28/21/213001} {\bibfield  {journal} {\bibinfo  {journal} {J. Phys.}\ }\textbf {\bibinfo {volume} {28}},\ \bibinfo {pages} {213001} (\bibinfo {year} {2016})}\BibitemShut {NoStop}%
\bibitem [{\citenamefont {Chen}\ \emph {et~al.}(2025)\citenamefont {Chen}, \citenamefont {Huang}, \citenamefont {Sun}, \citenamefont {Jie}, \citenamefont {Zhou}, \citenamefont {Hua}, \citenamefont {Xu}, \citenamefont {Wang}, \citenamefont {Guo}, \citenamefont {Zou}, \citenamefont {Sun},\ and\ \citenamefont {Zou}}]{chenOpenGRAPE}%
  \BibitemOpen
  \bibfield  {author} {\bibinfo {author} {\bibfnamefont {Z.-J.}\ \bibnamefont {Chen}}, \bibinfo {author} {\bibfnamefont {H.}~\bibnamefont {Huang}}, \bibinfo {author} {\bibfnamefont {L.}~\bibnamefont {Sun}}, \bibinfo {author} {\bibfnamefont {Q.-X.}\ \bibnamefont {Jie}}, \bibinfo {author} {\bibfnamefont {J.}~\bibnamefont {Zhou}}, \bibinfo {author} {\bibfnamefont {Z.}~\bibnamefont {Hua}}, \bibinfo {author} {\bibfnamefont {Y.}~\bibnamefont {Xu}}, \bibinfo {author} {\bibfnamefont {W.}~\bibnamefont {Wang}}, \bibinfo {author} {\bibfnamefont {G.-C.}\ \bibnamefont {Guo}}, \bibinfo {author} {\bibfnamefont {C.-L.}\ \bibnamefont {Zou}}, \bibinfo {author} {\bibfnamefont {L.}~\bibnamefont {Sun}}, \ and\ \bibinfo {author} {\bibfnamefont {X.-B.}\ \bibnamefont {Zou}},\ }\bibfield  {title} {\enquote {\bibinfo {title} {Robust and optimal control of open quantum systems},}\ }\href {\doibase 10.1126/sciadv.adr0875} {\bibfield  {journal} {\bibinfo  {journal} {Science Advances}\ }\textbf {\bibinfo {volume} {11}},\ \bibinfo {pages}
  {eadr0875} (\bibinfo {year} {2025})}\BibitemShut {NoStop}%
\bibitem [{\citenamefont {Schulte-Herbrüggen}\ \emph {et~al.}(2011)\citenamefont {Schulte-Herbrüggen}, \citenamefont {Spörl}, \citenamefont {Khaneja},\ and\ \citenamefont {Glaser}}]{schulte-herbruggenOptimalControlGenerating2011}%
  \BibitemOpen
  \bibfield  {author} {\bibinfo {author} {\bibfnamefont {T.}~\bibnamefont {Schulte-Herbrüggen}}, \bibinfo {author} {\bibfnamefont {A.}~\bibnamefont {Spörl}}, \bibinfo {author} {\bibfnamefont {N.}~\bibnamefont {Khaneja}}, \ and\ \bibinfo {author} {\bibfnamefont {S.~J.}\ \bibnamefont {Glaser}},\ }\bibfield  {title} {\enquote {\bibinfo {title} {Optimal control for generating quantum gates in open dissipative systems},}\ }\href {\doibase 10.1088/0953-4075/44/15/154013} {\bibfield  {journal} {\bibinfo  {journal} {J. Phys. B: At. Mol. Opt. Phys.}\ }\textbf {\bibinfo {volume} {44}},\ \bibinfo {pages} {154013} (\bibinfo {year} {2011})}\BibitemShut {NoStop}%
\bibitem [{\citenamefont {Boutin}\ \emph {et~al.}(2017)\citenamefont {Boutin}, \citenamefont {Andersen}, \citenamefont {Venkatraman}, \citenamefont {Ferris},\ and\ \citenamefont {Blais}}]{boutinResonatorResetCircuit2017a}%
  \BibitemOpen
  \bibfield  {author} {\bibinfo {author} {\bibfnamefont {S.}~\bibnamefont {Boutin}}, \bibinfo {author} {\bibfnamefont {C.~K.}\ \bibnamefont {Andersen}}, \bibinfo {author} {\bibfnamefont {J.}~\bibnamefont {Venkatraman}}, \bibinfo {author} {\bibfnamefont {A.~J.}\ \bibnamefont {Ferris}}, \ and\ \bibinfo {author} {\bibfnamefont {A.}~\bibnamefont {Blais}},\ }\bibfield  {title} {\enquote {\bibinfo {title} {Resonator reset in circuit qed by optimal control for large open quantum systems},}\ }\href {\doibase 10.1103/PhysRevA.96.042315} {\bibfield  {journal} {\bibinfo  {journal} {Phys. Rev. A}\ }\textbf {\bibinfo {volume} {96}},\ \bibinfo {pages} {042315} (\bibinfo {year} {2017})}\BibitemShut {NoStop}%
\bibitem [{\citenamefont {Abdelhafez}\ \emph {et~al.}(2019)\citenamefont {Abdelhafez}, \citenamefont {Schuster},\ and\ \citenamefont {Koch}}]{abdelhafez2019gradient-based}%
  \BibitemOpen
  \bibfield  {author} {\bibinfo {author} {\bibfnamefont {M.}~\bibnamefont {Abdelhafez}}, \bibinfo {author} {\bibfnamefont {D.~I.}\ \bibnamefont {Schuster}}, \ and\ \bibinfo {author} {\bibfnamefont {J.}~\bibnamefont {Koch}},\ }\bibfield  {title} {\enquote {\bibinfo {title} {{Gradient-based optimal control of open quantum systems using quantum trajectories and automatic differentiation}},}\ }\href {\doibase 10.1103/PhysRevA.99.052327} {\bibfield  {journal} {\bibinfo  {journal} {Phys. Rev. A}\ }\textbf {\bibinfo {volume} {99}},\ \bibinfo {pages} {52327} (\bibinfo {year} {2019})}\BibitemShut {NoStop}%
\bibitem [{\citenamefont {Goerz}\ \emph {et~al.}(2022)\citenamefont {Goerz}, \citenamefont {Carrasco},\ and\ \citenamefont {Malinovsky}}]{goerzQuantumOptimalControl2022}%
  \BibitemOpen
  \bibfield  {author} {\bibinfo {author} {\bibfnamefont {M.~H.}\ \bibnamefont {Goerz}}, \bibinfo {author} {\bibfnamefont {S.~C.}\ \bibnamefont {Carrasco}}, \ and\ \bibinfo {author} {\bibfnamefont {V.~S.}\ \bibnamefont {Malinovsky}},\ }\bibfield  {title} {\enquote {\bibinfo {title} {Quantum {{Optimal Control}} via {{Semi-Automatic Differentiation}}},}\ }\href {\doibase 10.22331/q-2022-12-07-871} {\bibfield  {journal} {\bibinfo  {journal} {Quantum}\ }\textbf {\bibinfo {volume} {6}},\ \bibinfo {pages} {871} (\bibinfo {year} {2022})}\BibitemShut {NoStop}%
\bibitem [{\citenamefont {Wu}\ \emph {et~al.}(2019)\citenamefont {Wu}, \citenamefont {Ding}, \citenamefont {Dong},\ and\ \citenamefont {Wang}}]{wuLearningRobustHighprecision2019}%
  \BibitemOpen
  \bibfield  {author} {\bibinfo {author} {\bibfnamefont {R.-B.}\ \bibnamefont {Wu}}, \bibinfo {author} {\bibfnamefont {H.}~\bibnamefont {Ding}}, \bibinfo {author} {\bibfnamefont {D.}~\bibnamefont {Dong}}, \ and\ \bibinfo {author} {\bibfnamefont {X.}~\bibnamefont {Wang}},\ }\bibfield  {title} {\enquote {\bibinfo {title} {Learning robust and high-precision quantum controls},}\ }\href {\doibase 10.1103/PhysRevA.99.042327} {\bibfield  {journal} {\bibinfo  {journal} {Phys. Rev. A}\ }\textbf {\bibinfo {volume} {99}},\ \bibinfo {pages} {042327} (\bibinfo {year} {2019})}\BibitemShut {NoStop}%
\bibitem [{\citenamefont {Chao}\ and\ \citenamefont {Reichardt}(2018)}]{chaoflagqubit2018}%
  \BibitemOpen
  \bibfield  {author} {\bibinfo {author} {\bibfnamefont {R.}~\bibnamefont {Chao}}\ and\ \bibinfo {author} {\bibfnamefont {B.~W.}\ \bibnamefont {Reichardt}},\ }\bibfield  {title} {\enquote {\bibinfo {title} {Quantum {{Error Correction}} with {{Only Two Extra Qubits}}},}\ }\href {\doibase 10.1103/PhysRevLett.121.050502} {\bibfield  {journal} {\bibinfo  {journal} {Phys. Rev. Lett.}\ }\textbf {\bibinfo {volume} {121}},\ \bibinfo {pages} {050502} (\bibinfo {year} {2018})}\BibitemShut {NoStop}%
\bibitem [{\citenamefont {Chamberland}\ and\ \citenamefont {Beverland}(2018)}]{chamberlandFlagFaulttolerantError2018}%
  \BibitemOpen
  \bibfield  {author} {\bibinfo {author} {\bibfnamefont {C.}~\bibnamefont {Chamberland}}\ and\ \bibinfo {author} {\bibfnamefont {M.~E.}\ \bibnamefont {Beverland}},\ }\bibfield  {title} {\enquote {\bibinfo {title} {Flag fault-tolerant error correction with arbitrary distance codes},}\ }\href {\doibase 10.22331/q-2018-02-08-53} {\bibfield  {journal} {\bibinfo  {journal} {Quantum}\ }\textbf {\bibinfo {volume} {2}},\ \bibinfo {pages} {53} (\bibinfo {year} {2018})}\BibitemShut {NoStop}%
\bibitem [{\citenamefont {Grassl}\ \emph {et~al.}(1997)\citenamefont {Grassl}, \citenamefont {Beth},\ and\ \citenamefont {Pellizzari}}]{grasslCodesQuantumErasure1997}%
  \BibitemOpen
  \bibfield  {author} {\bibinfo {author} {\bibfnamefont {M.}~\bibnamefont {Grassl}}, \bibinfo {author} {\bibfnamefont {{\relax Th}.}~\bibnamefont {Beth}}, \ and\ \bibinfo {author} {\bibfnamefont {T.}~\bibnamefont {Pellizzari}},\ }\bibfield  {title} {\enquote {\bibinfo {title} {Codes for the quantum erasure channel},}\ }\href {\doibase 10.1103/PhysRevA.56.33} {\bibfield  {journal} {\bibinfo  {journal} {Phys. Rev. A}\ }\textbf {\bibinfo {volume} {56}},\ \bibinfo {pages} {33} (\bibinfo {year} {1997})}\BibitemShut {NoStop}%
\bibitem [{\citenamefont {Fowler}\ \emph {et~al.}(2012)\citenamefont {Fowler}, \citenamefont {Mariantoni}, \citenamefont {Martinis},\ and\ \citenamefont {Cleland}}]{fowlerSurfaceCodesPractical2012}%
  \BibitemOpen
  \bibfield  {author} {\bibinfo {author} {\bibfnamefont {A.~G.}\ \bibnamefont {Fowler}}, \bibinfo {author} {\bibfnamefont {M.}~\bibnamefont {Mariantoni}}, \bibinfo {author} {\bibfnamefont {J.~M.}\ \bibnamefont {Martinis}}, \ and\ \bibinfo {author} {\bibfnamefont {A.~N.}\ \bibnamefont {Cleland}},\ }\bibfield  {title} {\enquote {\bibinfo {title} {Surface codes: {{Towards}} practical large-scale quantum computation},}\ }\href {\doibase 10.1103/PhysRevA.86.032324} {\bibfield  {journal} {\bibinfo  {journal} {Phys. Rev. A}\ }\textbf {\bibinfo {volume} {86}},\ \bibinfo {pages} {032324} (\bibinfo {year} {2012})}\BibitemShut {NoStop}%
\bibitem [{\citenamefont {Stace}\ \emph {et~al.}(2009)\citenamefont {Stace}, \citenamefont {Barrett},\ and\ \citenamefont {Doherty}}]{erasureerror-ThresholdsTopologicalCodes2009}%
  \BibitemOpen
  \bibfield  {author} {\bibinfo {author} {\bibfnamefont {T.~M.}\ \bibnamefont {Stace}}, \bibinfo {author} {\bibfnamefont {S.~D.}\ \bibnamefont {Barrett}}, \ and\ \bibinfo {author} {\bibfnamefont {A.~C.}\ \bibnamefont {Doherty}},\ }\bibfield  {title} {\enquote {\bibinfo {title} {Thresholds for {{Topological Codes}} in the {{Presence}} of {{Loss}}},}\ }\href {\doibase 10.1103/PhysRevLett.102.200501} {\bibfield  {journal} {\bibinfo  {journal} {Phys. Rev. Lett.}\ }\textbf {\bibinfo {volume} {102}},\ \bibinfo {pages} {200501} (\bibinfo {year} {2009})}\BibitemShut {NoStop}%
\bibitem [{\citenamefont {Wu}\ \emph {et~al.}(2022)\citenamefont {Wu}, \citenamefont {Kolkowitz}, \citenamefont {Puri},\ and\ \citenamefont {Thompson}}]{wu2022erasure-Rydbergatomarrays}%
  \BibitemOpen
  \bibfield  {author} {\bibinfo {author} {\bibfnamefont {Y.}~\bibnamefont {Wu}}, \bibinfo {author} {\bibfnamefont {S.}~\bibnamefont {Kolkowitz}}, \bibinfo {author} {\bibfnamefont {S.}~\bibnamefont {Puri}}, \ and\ \bibinfo {author} {\bibfnamefont {J.~D.}\ \bibnamefont {Thompson}},\ }\bibfield  {title} {\enquote {\bibinfo {title} {Erasure conversion for fault-tolerant quantum computing in alkaline earth {{Rydberg}} atom arrays},}\ }\href {\doibase 10.1038/s41467-022-32094-6} {\bibfield  {journal} {\bibinfo  {journal} {Nat. Commun.}\ }\textbf {\bibinfo {volume} {13}},\ \bibinfo {pages} {4657} (\bibinfo {year} {2022})}\BibitemShut {NoStop}%
\bibitem [{\citenamefont {Kubica}\ \emph {et~al.}(2023)\citenamefont {Kubica}, \citenamefont {Haim}, \citenamefont {Vaknin}, \citenamefont {Levine}, \citenamefont {Brand{\~a}o},\ and\ \citenamefont {Retzker}}]{kubicaErasureQubitsOvercoming2023}%
  \BibitemOpen
  \bibfield  {author} {\bibinfo {author} {\bibfnamefont {A.}~\bibnamefont {Kubica}}, \bibinfo {author} {\bibfnamefont {A.}~\bibnamefont {Haim}}, \bibinfo {author} {\bibfnamefont {Y.}~\bibnamefont {Vaknin}}, \bibinfo {author} {\bibfnamefont {H.}~\bibnamefont {Levine}}, \bibinfo {author} {\bibfnamefont {F.}~\bibnamefont {Brand{\~a}o}}, \ and\ \bibinfo {author} {\bibfnamefont {A.}~\bibnamefont {Retzker}},\ }\bibfield  {title} {\enquote {\bibinfo {title} {Erasure {{Qubits}}: {{Overcoming}} the {{T}} 1 {{Limit}} in {{Superconducting Circuits}}},}\ }\href {\doibase 10.1103/PhysRevX.13.041022} {\bibfield  {journal} {\bibinfo  {journal} {Phys. Rev. X}\ }\textbf {\bibinfo {volume} {13}},\ \bibinfo {pages} {041022} (\bibinfo {year} {2023})}\BibitemShut {NoStop}%
\bibitem [{\citenamefont {Blais}\ \emph {et~al.}(2021)\citenamefont {Blais}, \citenamefont {Grimsmo}, \citenamefont {Girvin},\ and\ \citenamefont {Wallraff}}]{blaisCircuitQuantumElectrodynamics2021}%
  \BibitemOpen
  \bibfield  {author} {\bibinfo {author} {\bibfnamefont {A.}~\bibnamefont {Blais}}, \bibinfo {author} {\bibfnamefont {A.~L.}\ \bibnamefont {Grimsmo}}, \bibinfo {author} {\bibfnamefont {S.~M.}\ \bibnamefont {Girvin}}, \ and\ \bibinfo {author} {\bibfnamefont {A.}~\bibnamefont {Wallraff}},\ }\bibfield  {title} {\enquote {\bibinfo {title} {Circuit quantum electrodynamics},}\ }\href {\doibase 10.1103/RevModPhys.93.025005} {\bibfield  {journal} {\bibinfo  {journal} {Rev. Mod. Phys.}\ }\textbf {\bibinfo {volume} {93}},\ \bibinfo {pages} {025005} (\bibinfo {year} {2021})}\BibitemShut {NoStop}%
\bibitem [{\citenamefont {Vool}\ and\ \citenamefont {Devoret}(2017)}]{voolIntroductionQuantumElectromagnetic2017}%
  \BibitemOpen
  \bibfield  {author} {\bibinfo {author} {\bibfnamefont {U.}~\bibnamefont {Vool}}\ and\ \bibinfo {author} {\bibfnamefont {M.}~\bibnamefont {Devoret}},\ }\bibfield  {title} {\enquote {\bibinfo {title} {Introduction to quantum electromagnetic circuits},}\ }\href {\doibase 10.1002/cta.2359} {\bibfield  {journal} {\bibinfo  {journal} {International Journal of Circuit Theory and Applications}\ }\textbf {\bibinfo {volume} {45}},\ \bibinfo {pages} {897} (\bibinfo {year} {2017})}\BibitemShut {NoStop}%
\bibitem [{\citenamefont {Lindblad}(1976)}]{lindblad_generators_1976}%
  \BibitemOpen
  \bibfield  {author} {\bibinfo {author} {\bibfnamefont {G.}~\bibnamefont {Lindblad}},\ }\bibfield  {title} {{\selectlanguage {english}\enquote {\bibinfo {title} {On the generators of quantum dynamical semigroups},}\ }}\href {\doibase 10.1007/BF01608499} {\bibfield  {journal} {\bibinfo  {journal} {Comm. in Math. Phys.}\ }\textbf {\bibinfo {volume} {48}},\ \bibinfo {pages} {119} (\bibinfo {year} {1976})}\BibitemShut {NoStop}%
\bibitem [{\citenamefont {Gorini}\ \emph {et~al.}(1976)\citenamefont {Gorini}, \citenamefont {Kossakowski},\ and\ \citenamefont {Sudarshan}}]{goriniCompletelyPositiveDynamical1976}%
  \BibitemOpen
  \bibfield  {author} {\bibinfo {author} {\bibfnamefont {V.}~\bibnamefont {Gorini}}, \bibinfo {author} {\bibfnamefont {A.}~\bibnamefont {Kossakowski}}, \ and\ \bibinfo {author} {\bibfnamefont {E.~C.~G.}\ \bibnamefont {Sudarshan}},\ }\bibfield  {title} {\enquote {\bibinfo {title} {Completely positive dynamical semigroups of n -level systems},}\ }\href {\doibase 10.1063/1.522979} {\bibfield  {journal} {\bibinfo  {journal} {J. Math. Phys.}\ }\textbf {\bibinfo {volume} {17}},\ \bibinfo {pages} {821} (\bibinfo {year} {1976})}\BibitemShut {NoStop}%
\bibitem [{\citenamefont {Sch{\"a}fer}\ \emph {et~al.}(2021)\citenamefont {Sch{\"a}fer}, \citenamefont {Sekatski}, \citenamefont {Koppenh{\"o}fer}, \citenamefont {Bruder},\ and\ \citenamefont {Kloc}}]{schaferControlStochasticQuantum2021}%
  \BibitemOpen
  \bibfield  {author} {\bibinfo {author} {\bibfnamefont {F.}~\bibnamefont {Sch{\"a}fer}}, \bibinfo {author} {\bibfnamefont {P.}~\bibnamefont {Sekatski}}, \bibinfo {author} {\bibfnamefont {M.}~\bibnamefont {Koppenh{\"o}fer}}, \bibinfo {author} {\bibfnamefont {C.}~\bibnamefont {Bruder}}, \ and\ \bibinfo {author} {\bibfnamefont {M.}~\bibnamefont {Kloc}},\ }\bibfield  {title} {\enquote {\bibinfo {title} {Control of stochastic quantum dynamics by differentiable programming},}\ }\href {\doibase 10.1088/2632-2153/abec22} {\bibfield  {journal} {\bibinfo  {journal} {Machine Learning: Science and Technology}\ }\textbf {\bibinfo {volume} {2}},\ \bibinfo {pages} {035004} (\bibinfo {year} {2021})}\BibitemShut {NoStop}%
\bibitem [{\citenamefont {Propson}\ \emph {et~al.}(2022)\citenamefont {Propson}, \citenamefont {Jackson}, \citenamefont {Koch}, \citenamefont {Manchester},\ and\ \citenamefont {Schuster}}]{propsonRobustQuantumOptimal2022}%
  \BibitemOpen
  \bibfield  {author} {\bibinfo {author} {\bibfnamefont {T.}~\bibnamefont {Propson}}, \bibinfo {author} {\bibfnamefont {B.~E.}\ \bibnamefont {Jackson}}, \bibinfo {author} {\bibfnamefont {J.}~\bibnamefont {Koch}}, \bibinfo {author} {\bibfnamefont {Z.}~\bibnamefont {Manchester}}, \ and\ \bibinfo {author} {\bibfnamefont {D.~I.}\ \bibnamefont {Schuster}},\ }\bibfield  {title} {\enquote {\bibinfo {title} {Robust {{Quantum Optimal Control}} with {{Trajectory Optimization}}},}\ }\href {\doibase 10.1103/PhysRevApplied.17.014036} {\bibfield  {journal} {\bibinfo  {journal} {Physical Review Applied}\ }\textbf {\bibinfo {volume} {17}},\ \bibinfo {pages} {014036} (\bibinfo {year} {2022})}\BibitemShut {NoStop}%
\bibitem [{\citenamefont {Kraus}(1971)}]{kraus_general_1971}%
  \BibitemOpen
  \bibfield  {author} {\bibinfo {author} {\bibfnamefont {K.}~\bibnamefont {Kraus}},\ }\bibfield  {title} {{\selectlanguage {english}\enquote {\bibinfo {title} {General state changes in quantum theory},}\ }}\href {\doibase 10.1016/0003-4916(71)90108-4} {\bibfield  {journal} {\bibinfo  {journal} {Annals of Physics}\ }\textbf {\bibinfo {volume} {64}},\ \bibinfo {pages} {311} (\bibinfo {year} {1971})}\BibitemShut {NoStop}%
\bibitem [{\citenamefont {Chen}\ \emph {et~al.}(2026)\citenamefont {Chen}, \citenamefont {Cai}, \citenamefont {Xie}, \citenamefont {Jie}, \citenamefont {Zou}, \citenamefont {Guo}, \citenamefont {Sun},\ and\ \citenamefont {Zou}}]{chen2026faulttolerantpreparationarbitrarylogical}%
  \BibitemOpen
  \bibfield  {author} {\bibinfo {author} {\bibfnamefont {Z.-J.}\ \bibnamefont {Chen}}, \bibinfo {author} {\bibfnamefont {W.}~\bibnamefont {Cai}}, \bibinfo {author} {\bibfnamefont {L.-X.}\ \bibnamefont {Xie}}, \bibinfo {author} {\bibfnamefont {Q.-X.}\ \bibnamefont {Jie}}, \bibinfo {author} {\bibfnamefont {X.-B.}\ \bibnamefont {Zou}}, \bibinfo {author} {\bibfnamefont {G.-C.}\ \bibnamefont {Guo}}, \bibinfo {author} {\bibfnamefont {L.}~\bibnamefont {Sun}}, \ and\ \bibinfo {author} {\bibfnamefont {C.-L.}\ \bibnamefont {Zou}},\ }\bibfield  {title} {\enquote {\bibinfo {title} {Fault-tolerant preparation of arbitrary logical states in the cat code},}\ }\href {\doibase 10.48550/arXiv.2602.17438} {\bibfield  {journal} {\bibinfo  {journal} {arXiv:2602.17438}\ } (\bibinfo {year} {2026}),\ 10.48550/arXiv.2602.17438}\BibitemShut {NoStop}%
\bibitem [{\citenamefont {Paul}(1990)}]{paulElectromagneticTrapsCharged1990}%
  \BibitemOpen
  \bibfield  {author} {\bibinfo {author} {\bibfnamefont {W.}~\bibnamefont {Paul}},\ }\bibfield  {title} {\enquote {\bibinfo {title} {Electromagnetic traps for charged and neutral particles},}\ }\href {\doibase 10.1103/RevModPhys.62.531} {\bibfield  {journal} {\bibinfo  {journal} {Reviews of Modern Physics}\ }\textbf {\bibinfo {volume} {62}},\ \bibinfo {pages} {531} (\bibinfo {year} {1990})}\BibitemShut {NoStop}%
\bibitem [{\citenamefont {Leibfried}\ \emph {et~al.}(2003)\citenamefont {Leibfried}, \citenamefont {Blatt}, \citenamefont {Monroe},\ and\ \citenamefont {Wineland}}]{leibfriedSingletrappedions2003}%
  \BibitemOpen
  \bibfield  {author} {\bibinfo {author} {\bibfnamefont {D.}~\bibnamefont {Leibfried}}, \bibinfo {author} {\bibfnamefont {R.}~\bibnamefont {Blatt}}, \bibinfo {author} {\bibfnamefont {C.}~\bibnamefont {Monroe}}, \ and\ \bibinfo {author} {\bibfnamefont {D.}~\bibnamefont {Wineland}},\ }\bibfield  {title} {\enquote {\bibinfo {title} {Quantum dynamics of single trapped ions},}\ }\href {\doibase 10.1103/RevModPhys.75.281} {\bibfield  {journal} {\bibinfo  {journal} {Reviews of Modern Physics}\ }\textbf {\bibinfo {volume} {75}},\ \bibinfo {pages} {281} (\bibinfo {year} {2003})}\BibitemShut {NoStop}%
\bibitem [{\citenamefont {Aliferis}\ and\ \citenamefont {Preskill}(2008)}]{aliferisFTagainstbiasednoise2008}%
  \BibitemOpen
  \bibfield  {author} {\bibinfo {author} {\bibfnamefont {P.}~\bibnamefont {Aliferis}}\ and\ \bibinfo {author} {\bibfnamefont {J.}~\bibnamefont {Preskill}},\ }\bibfield  {title} {\enquote {\bibinfo {title} {Fault-tolerant quantum computation against biased noise},}\ }\href {\doibase 10.1103/PhysRevA.78.052331} {\bibfield  {journal} {\bibinfo  {journal} {Physical Review A}\ }\textbf {\bibinfo {volume} {78}},\ \bibinfo {pages} {052331} (\bibinfo {year} {2008})}\BibitemShut {NoStop}%
\bibitem [{\citenamefont {Tuckett}\ \emph {et~al.}(2018)\citenamefont {Tuckett}, \citenamefont {Bartlett},\ and\ \citenamefont {Flammia}}]{tuckettbiased2018}%
  \BibitemOpen
  \bibfield  {author} {\bibinfo {author} {\bibfnamefont {D.~K.}\ \bibnamefont {Tuckett}}, \bibinfo {author} {\bibfnamefont {S.~D.}\ \bibnamefont {Bartlett}}, \ and\ \bibinfo {author} {\bibfnamefont {S.~T.}\ \bibnamefont {Flammia}},\ }\bibfield  {title} {\enquote {\bibinfo {title} {Ultrahigh {{Error Threshold}} for {{Surface Codes}} with {{Biased Noise}}},}\ }\href {\doibase 10.1103/PhysRevLett.120.050505} {\bibfield  {journal} {\bibinfo  {journal} {Physical Review Letters}\ }\textbf {\bibinfo {volume} {120}},\ \bibinfo {pages} {050505} (\bibinfo {year} {2018})}\BibitemShut {NoStop}%
\bibitem [{\citenamefont {Bonilla~Ataides}\ \emph {et~al.}(2021)\citenamefont {Bonilla~Ataides}, \citenamefont {Tuckett}, \citenamefont {Bartlett}, \citenamefont {Flammia},\ and\ \citenamefont {Brown}}]{bonillaataidesXZZXSurfaceCode2021}%
  \BibitemOpen
  \bibfield  {author} {\bibinfo {author} {\bibfnamefont {J.~P.}\ \bibnamefont {Bonilla~Ataides}}, \bibinfo {author} {\bibfnamefont {D.~K.}\ \bibnamefont {Tuckett}}, \bibinfo {author} {\bibfnamefont {S.~D.}\ \bibnamefont {Bartlett}}, \bibinfo {author} {\bibfnamefont {S.~T.}\ \bibnamefont {Flammia}}, \ and\ \bibinfo {author} {\bibfnamefont {B.~J.}\ \bibnamefont {Brown}},\ }\bibfield  {title} {\enquote {\bibinfo {title} {The {{XZZX}} surface code},}\ }\href {\doibase 10.1038/s41467-021-22274-1} {\bibfield  {journal} {\bibinfo  {journal} {Nature Communications}\ }\textbf {\bibinfo {volume} {12}},\ \bibinfo {pages} {2172} (\bibinfo {year} {2021})}\BibitemShut {NoStop}%
\end{thebibliography}%

\end{document}